# Nanoparticle aggregation controlled by desalting kinetics

J. Fresnais[1], C. Lavelle[2] and J.-F. Berret[1,@]

[1] *Matière et Systèmes Complexes, UMR 7057 CNRS Université Denis Diderot Paris-VII, Bâtiment Condorcet, F-75205 Paris, France*
[2] *Interdisciplinary Research Institute, USR 3078 CNRS Université Lille I, F-59655 Villeneuve d'Ascq, France*

**Abstract** : We report the formation of stable nanoparticle-polymer clusters obtained by electrostatic complexation. The nanoparticles placed under scrutiny are nanoceria ($CeO_2$) coated by short poly(acrylic acid) moieties, whereas the polymers are cationic-neutral block copolymers. The cluster formation was monitored using different formulation pathways, including direct mixing, dialysis, dilution and quenching. In the first process, the hybrids were obtained by mixing stock solutions of polymers and nanoparticles. Dialysis and dilution were based on controlled desalting kinetics according to methods developed in molecular biology. The fourth process consisted in a rapid dilution of the salted dispersions and as such it was regarded as a quench of the cluster kinetics. We have found that one key parameter that controls the kinetics of formation of electrostatic clusters is the rate $dI_S/dt$ at which the salt is removed from the solution, where $I_S$ denotes the ionic strength. With decreasing $I_S$, the electrostatically screened polymers and nanoparticles system undergo an abrupt transition between an unassociated and a clustered state. By tuning the desalting kinetics, the size of the clusters was varied from 100 nm to over 1 μm. At low ionic strength, the clusters were found to be kinetically frozen. It is proposed that the onset of aggregation is driven by the desorption-adsorption transition of the polymers onto the surfaces of the particles.

## I – Introduction

The electrostatic complexation between oppositely charged species in aqueous solutions has attracted much attention during the last years because it is thought to be at the origin of many fundamental assembly mechanisms.[1-5] These mechanisms were also recognized as essential in several applications, including water and waste treatments,[4] transfection vectors and targeting in biology[6-8] and design of new colloidal structures in material science.[9,10] In general, the attractive interactions between oppositely charged colloids or macromolecules are strong, and the direct mixing of solutions containing such entities yields a phase separation. This is the case e.g. for anionic polyelectrolytes and cationic surfactants, for which micellar coacervate and liquid crystalline phases have been observed.[11-13] Means to control the electrostatically-driven attractions and to preserve the colloidal stability were developed using copolymers, and in particular polyelectrolyte-neutral block copolymers.[5,8,14] These fully hydrosoluble macromolecules were found to co-assemble spontaneously with different types of systems, such as surfactants,[15-17] polymers[18,19] and proteins,[20] yielding core-shell structures. As a result of the co-assembly, the cores of the aggregates were described as a dense coacervate microphase comprising the oppositely charged species, and surrounded by a neutral corona made from the neutral blocks.

So far, electrostatic aggregates were obtained according to different formulation modes : direct mixing of the stock dispersions, or titration of one solution with another. Although it was recognized that coacervate micelles could be kinetically frozen,[21] no systematic investigation of their sizes and morphologies as a function of the formulation was undertaken. The electrostatic complexation between spherical nanocolloids of sizes 10 nm or less and copolymers have shown a diversity of results. With surfactants, the core-shell structures exhibited spherical symmetry and monodisperse cores,[17] whereas with nanoparticles (either cerium or iron oxide),[22] elongated and branched aggregates were systematically found out. It was suggested that the variety of aggregate morphologies seen with oxide based hybrids could stem from an uncontrolled kinetics of association between the oppositely charged species. For sake of completeness, it should mentioned that alternative approaches of hybrid formation based on *in situ* synthesis of particles in presence of polymers[23-28] or encapsulation within hydrophobic cores[29-33] were also reported.

In the present paper, we have considered the anionically charged nanoceria[22,34] and cationic-neutral block copolymers[17,35] and propose new protocols for bringing oppositely charged species together. In terms of applications, cerium oxide remains an essential ingredient in catalysis, coating and biomedicine.[36-39] For chemical-mechanical polishing[40], particulates of sizes comprised between 50 nm – 1 μm are required, but remain so far difficult to generate by soft chemistry routes. The proposed clustering approaches offer opportunities to elaborate stable mineral colloids devoted to these applications. The protocols typically were inspired by molecular biology techniques developed for the *in vitro* reconstitutions of chromatin, the DNA / histones macromolecular substance that forms the chromosomes of our cells.[41] The protocols consisted of two steps.[42,43] The first step was based on the screening of the electrostatic interactions by bringing the dispersions to 1 M of salt. In the second step, the salt was removed progressively by dialysis or by dilution. In the present work, four different formulation techniques were compared : **direct mixing**, **dialysis**, **dilution** and **quenching**. We have found that one of the key parameters that controls the kinetics of formation of the electrostatic clusters was the rate at which the salt was removed from the dispersion. More importantly, it was established that with decreasing ionic strength the electrostatically screened polymer and nanoparticle system underwent an abrupt transition between an unassociated and a cluster state, and that this critical ionic strength (around 0.4 M) did not depend on the formulation process. It was finally suggested that the cluster formation bears strong similarities with the elementary processes found in the kinetics of first order phase transitions.

## II – Methods

### II.1 - Particles and polymers

Complexation schemes were realized using cerium oxide nanoparticles and oppositely charged block copolymers. Nanoceria was synthesized by thermohydrolysis of cerium(IV) nitrate salt in acidic condition and high temperature.[22,34] The particles consisted of isotropic agglomerates of 2 - 5 crystallites with typical size 2 nm and faceted morphologies. The average diameter of the bare particles was 7.0 nm with a polydispersity (ratio between the standard deviation and the average diameter) of 0.15. The particles were coated by poly(acrylic acid) oligomers with molecular weight 2000 g mol$^{-1}$, using the precipitation-redispersion process described previously.[34] The hydrodynamic sizes found in $CeO_2$-$PAA_{2K}$ dispersions was $D_H$ = 13 nm, i.e. 6 nm above that of the bare particles. This increase was assigned to the presence of a 3 nm-$PAA_{2K}$ brush around the particles. In the conditions at which the complexation was realized (neutral pH and no added salt), the particles were stabilized by electrostatic interactions mediated by the ionized carboxylate groups of the $PAA_{2K}$-chains. The nanoparticles were thus considered as being anionically charged.

The cationic-neutral block copolymer used in the complexation was poly(trimethylammonium ethylacrylate)-b-poly(acrylamide) with molecular weight 11000 g mol$^{-1}$ (41 monomers) for the charged block and 30000 g mol$^{-1}$ (420 monomers) for the neutral block. The diblock copolymers, abbreviated $PTEA_{11K}$-b-$PAM_{30K}$ in the following were synthesized by controlled radical polymerization according to MADIX technology[44,45] and we refer to earlier work for the chemical formulae of the monomers and detailed characterization.[17,35] In dilute solution and at neutral pH, the chains were well dispersed and in the state of unimers, with an hydrodynamic diameter of 11 nm. The polydispersity index was determined by size exclusion chromatography at 1.6. Photographs of the initial nanoparticle and polymer solutions are shown in Fig. 1 (VIAL **a** and VIAL **b**, respectively).

### II.2 - Sample Preparation

Nanoparticles clusters were prepared according to four different methods. The first method, called **direct mixing**, utilized stock polymer and nanoparticle solutions prepared without added salt. The three other protocols, **dialysis**, **dilution** and **quenching** were based on a principle of desalting processes, starting all runs at the initial ionic strength $I_S^i$ = 1 M of ammonium chloride ($NH_4Cl$).[46] These protocols differentiate themselves by the kinetics of the removal of the salt, and by different jumps in ionic strength performed. The ionic strength was defined as[47]:

$$I_S = \frac{1}{2}\sum_i c_i Z_i^2 \qquad (1)$$

where $c_i$ and $Z_i$ denote the concentration and valency of the ionic atomic species in solution.

**Direct mixing** : Polymer-nanoparticle complexes were obtained by mixing stock solutions prepared at the same weight concentration (c ~ 0.1 wt. %) and same pH (pH 8). The mixing of the two initial solutions was characterized by the ratio X = $V_{Part}/V_{Pol}$, where $V_{Part}$ and $V_{Pol}$ were the volumes of the particle and polymer solutions, respectively. This procedure was preferred to titration experiments, because it allowed to explore a broad range in mixing ratios (X = $10^{-2}$ – 100) and simultaneously to keep the total concentration in the dilute regime.[35] As far as the kinetics is concerned, the formation of the aggregates occurred very rapidly on mixing, i.e. within a time scale inferior to one second. In the ranges investigated, the dispersions resulting from direct mixing were fully reproducible.

|   | pathways | $I_S^i$ (M) | $I_S^f$ (M) | $\Delta I_S = I_S^i - I_S^f$ (M) | $dI_S/dt$ (M s$^{-1}$) | $D_H$ nm |
|---|---|---|---|---|---|---|
| 1 | direct mixing | $5\times10^{-3}$ | $5\times10^{-3}$ | 0 | .. | 120 |
| 2 | dialysis | 1 | $5\times10^{-3}$ | ~ 1 | $10^{-4}$ | 300 - 500 |
| 3 | Dilution | 1 | $5\times10^{-2}$ | 0.95 | $10^{-4}$ - 1 | 100 - 500 |
| 4 | Quenching | 1 | $5\times10^{-2}$ – 1 | 0 – 0.95 | ~ 1 | 100 |

**Table I** : List of the different mixing and desalting configurations tested in this work. $I_S^i$ and $I_S^f$ denote the initial and final ionic strengths, respectively, whereas $dI_S/dt$ is the rate at which the ionic strength was lowered. In the direct mixing process, the ionic strength was estimated from the free counterions present in the solution. The $D_H$-value given in the first line of the Table was that found at the maximum scattering intensity (i.e. at $X_P$ = 1.6). With the present terminology, rapid dilution and quenching at low ionic strength are identical.

**Dialysis** : Mixtures of $PAA_{2K}$-coated $CeO_2$ nanoparticles and $PTEA_{11K}$-b-$PAM_{30K}$ copolymers in 1 M $NH_4Cl$ were dialyzed against deionized water at pH 7 using a Slide-A-Lyser cassette with a 10 kD cutoff membrane.[46] The volume of the dialysis bath was 100 times larger than that of the samples. The electrical conductivity was measured during the ion exchange and served to monitor the desalting kinetics. In the condition described here, the whole process reached a stationary and final state within 2 hours. The dialysis experiment between the initial $I_S^i$ and final $I_S^f$ ionic strengths was characterized by an average rate of ionic strength change $dI_S/dt$ ~ -$10^{-4}$ M s$^{-1}$ (see Table I for details). Although in all the desalting experiments the rate of change was negative, in the following the negative sign will be omitted for simplicity. Note that with dialysis, the nanoparticles concentration remained practically constant (c ~ 0.1 wt. %).

**Dilution** : In the dilution process, deionized water was added to mixtures of $PAA_{2K}$-coated $CeO_2$ nanoparticles and $PTEA_{11K}$-b-$PAM_{30K}$ copolymer stepwise, changing $I_S$ from 1 M to $5\times10^{-2}$ M. In this process, the overall concentration was decreased by a factor 20. Since the aggregates formed by dilution are large compared to the unassociated polymer and particles, the measurements of their hydrodynamic properties up to the lowest ionic strength could be easily realized. In a typical experiment, the rate of change in ionic strength was taken to be that at the transition, namely :

$$\frac{dI_S}{dt} = -I_S^{c\,2} \frac{Q}{I_S^i v_0} \quad (2)$$

where Q was the flow rate and $v_0$ the initial volume. Eq. 2 was derived from the expression of the time dependent ionic strength : $I_S(t) = I_S^i v_0/(v_0 + Qt)$. By varying the different parameters in Eq. 2, $dI_S/dt$ was changed from $10^{-4}$ M s$^{-1}$ to 1 M s$^{-1}$. The critical transition rate noted $I_S^c$ is defined in Section III.

*Quenching* : Quenching corresponded to a rapid dilution at a rate $dI_S/dt \sim 1$ M s$^{-1}$. Experiments were run in solvents of different ionic strength, *i.e.* for $I_S^f$ comprised between $5\times10^{-3}$ M and 1 M. For quenching tests performed using de-ionized water, Eq. 2 was applied to estimate the rate of change in ionic strength. Once the quenching of the initial 1 M solution was realized, the samples were let free to equilibrate for several days before measurements were undertaken. Table I summarizes the experimental conditions of the four procedures in terms of amplitudes and rate of changes of the ionic strength. The photographs in Fig. 1c shows a typical dispersion obtained by direct mixing, whereas Figs. 1d - 1f display mixed solutions obtained from dialysis, dilution and quenching, respectively. The four vials in the lower panel of Fig. 1 exhibit distinctive light scattering properties that are described in the Results and Discussion Section.

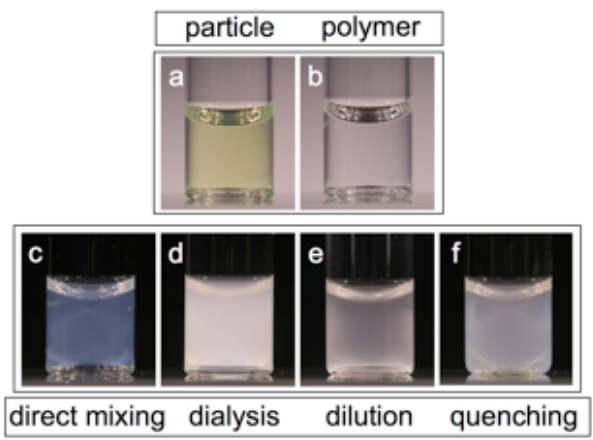

*Figure 1* : Photographs of the nanoparticle (VIAL *a*), polymer (VIAL *b*), and cluster dispersions (VIALS *c-f*). Except for the nanoparticles, the concentrations are c = 0.2 wt. %. The concentration in VIAL *a* is 20 wt. % so as to show the absorption property of nanoceria in the visible range. The vials in the lower panel were obtained via direct mixing (VIAL *c*), dialysis (VIAL *d*), dilution (VIAL *e*) and quenching (VIAL *f*).

II.3 - Cryogenic transmission electron microscopy and light scattering

Cryo-transmission electron microscopy (cryo-TEM) was performed on dilute solutions (c ~ 0.1 wt. %). For the experiments, a drop of the solution was put on a TEM-grid covered by a 100 nm-thick polymer perforated membrane. The drop was blotted with filter paper and the grid was quenched rapidly in liquid ethane in order to avoid the crystallization of the aqueous phase. The membrane was then transferred into the vacuum column of a TEM-microscope (JEOL 1200 EX operating at 120 kV) where it was maintained at liquid nitrogen temperature. The magnification for the cryo-TEM experiments was selected at 40 000×. Transmission electron microscopy (TEM) was carried out on a Jeol-100 CX microscope at the SIARE facility of University Pierre et Marie Curie (Paris 6).

Static and dynamic light scattering were monitored on a Brookhaven spectrometer (BI-9000AT autocorrelator) and on a NanoZS (Malvern Instrument) for measurements of the Rayleigh ratio $\mathcal{R}(q,c)$ and of the collective diffusion constant D(c). The Rayleigh ratio was obtained from the scattered intensity I(q,c) measured at the wave-vector q according to[48] :

$$\mathcal{R}(q,c) = \mathcal{R}_{Std} \frac{I(q,c) - I_{Water}}{I_{Tol}} \left(\frac{n}{n_{Tol}}\right)^2 \quad (3)$$

Here, $\mathcal{R}_{Std}$ and $n_{Tol}$ are the standard Rayleigh ratio and refractive index of toluene, $I_{Water}$ and $I_{Tol}$ the intensities measured for the solvent and for the toluene in the same scattering configuration and $q q = 4\pi n/\lambda \sin(\theta/2)$ (n being the refractive index of the solution and $\theta$ the scattering angle). In this study, the Rayleigh ratio $\mathcal{R}(q,c)$ was measured as a function of the mixing ratio X and for the different desalting kinetics. With the Brookhaven spectrometer, the scattering angle was $\theta = 90°$, whereas for the NanoZS it was $\theta = 173°$, corresponding to wave-vectors $q = 1.87\times10^{-3}$ Å$^{-1}$ and $q = 2.64\times10^{-3}$ Å$^{-1}$, respectively. In quasi-elastic light scattering, the collective diffusion coefficient $D_0$ was measured in the dilute concentration range (c = 0.1 wt. %). The hydrodynamic diameter of the colloids was calculated according to the Stokes-Einstein relation, $D_H = k_B T/3\pi\eta D_0$, where $k_B$ is the Boltzmann constant, T the temperature (T = 298 K) and $\eta$ the solvent viscosity ($0.89\times10^{-3}$ Pa s). For values of the hydrodynamic diameter such as $qD_H > 1$, the Stokes-Einstein relationship was still used, but the hydrodynamic data were associated with larger uncertainties. As shown by Dhont[49] in this range, the collective diffusion coefficient depended on the wave-vector and on the polydispersity, with deviations up to 10 % with respect to $D_0$ (for polydispersity 0.1). The autocorrelation functions of the scattered light were interpreted using both the method of cumulants and the CONTIN fitting procedure provided by the instrument software.

## III – Results and discussion

### III.1 - Direct Mixing

Fig. 2a displays the normalized Rayleigh ratios $\mathcal{R}(X)/\mathcal{R}_\infty$ obtained at $q = 2.64\times10^{-3}$ Å$^{-1}$ for CeO$_2$-PAA$_{2K}$ complexed with PTEA$_{11K}$-b-PAM$_{30K}$ copolymers, for X ranging from $10^{-2}$ to 100 (c = 0.2 wt. %, T = 25° C). The polymer (X = 0) and nanoparticle (X = ∞) stock solutions have been set at $X = 10^{-3}$ and X = 1000 for convenience. In the figure, the scattered intensity was found to increase steadily with X, to pass through a sharp maximum at $X_P = 1.6 \pm 0.2$ and then to decrease to 1. The normalization factor in Fig. 2a, $\mathcal{R}_\infty = 3.2\times10^{-4}$ cm$^{-1}$ represents the intensity of the coated nanoparticles. By analogy with a recent work on complexation of similar systems, the data in Fig. 2a were compared with the predictions of a *stoichiometric model*.

This model assumes that the mixed aggregates are formed at a fixed polymer-to-nanoparticle ratio (noted $X_P$) and that the stoichiometry of the mixed aggregates is controlled by the structural charges present.

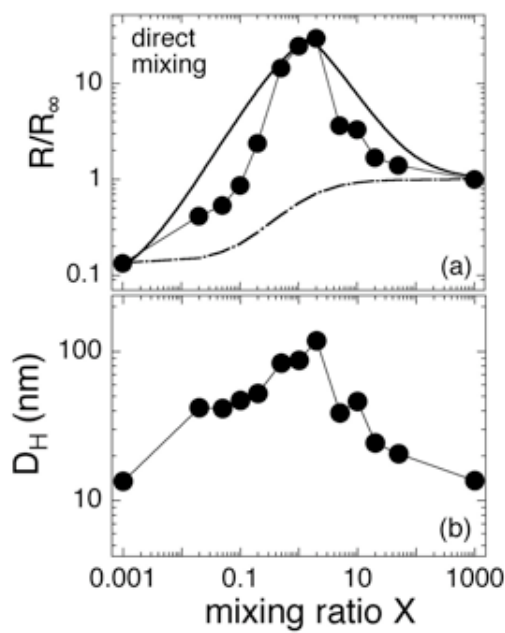

*Figure 2* : a) Normalized Rayleigh ratios $\mathcal{R}(X)/\mathcal{R}_\infty$ obtained at $q = 2.65 \times 10^{-3}$ Å$^{-1}$ for CeO$_2$-PAA$_{2K}$ nanoparticles complexed with PTEA$_{11K}$-b-PAM$_{30K}$ block copolymers (direct mixing). The total concentration is c = 0.2 wt. % and temperature T = 25° C. The solid line results from best fit calculations using a stoichiometric association model (Eq. A-5). The dashed line represents the scattering intensity for unassociated particles and polymers. The thin lines between the data points are guides for the eyes. b) Hydrodynamic diameter $D_H$ as function of the mixing ratio X for the same system.

In this model, the maximum scattering corresponds to a state where all the polymers and all particles co-assemble, yielding the largest number density of complexes for a given concentration. The stoichiometric model is described in detail in Appendix. We have found that 8 polymers per particle and 30 particles per aggregate were necessary to account for the data. The result of the fitting is shown in Fig. 2a as a solid line. The agreement between the model and the data is reasonable. There, the position and amplitude of the scattering peaks are well accounted for by the predictions of Eq. A-5. In the figure is also displayed for comparison the scattering intensity corresponding to the state where particles and polymers remain unaggregated (dashed line, Eq. A-7).

Dynamic light scattering performed on the same samples revealed the presence of a unique diffusive relaxation mode over a wide range of X. Fig. 2b displays the evolution of the hydrodynamic diameters $D_H$ determined by the Stokes-Einstein relation. For X > 0.01, the hydrodynamic diameter is much larger than those of the polymers and of the nanoparticles. It ranged between 40 nm and 120 nm, with a maximum coinciding with the maximum scattering intensity (VIAL **c** in Fig. 1). For X > 10, the diameter decreased toward that of the coated nanoparticles. The polydispersity indices obtained from the cumulant analysis were found for these systems in the range 0.10 – 0.25.[20,21] Except at very large X, the data displayed in Figs. 2 bear strong similarities with those reported recently, using citrate coated particles and cationic-neutral block copolymers.[50] Electrophoretic mobility measurements have shown that the mixed aggregates are neutral or close to neutrality. Values in the range (-0.3…-0.03)×10$^{-4}$ cm$^2$ V$^{-1}$ s$^{-1}$ were found, *i.e.* 10 times lower than the mobilities of the PAA$_{2K}$-coated particles.[35] Note finally that these aggregates exhibit a better colloidal stability with respect to concentration, pH and ionic strength changes, as compared to the single nanoparticles.

Fig. 3 exhibits a cryo-TEM image of the polymers-nanoparticles aggregates. For contrast reasons, only the inorganic cores are visible with this technique. The thickness of the polymer corona (PAM) around the cores was estimated by light scattering at ~ 20 nm.[50] The aggregates were found to be polydisperse, with sizes comprised between 20 and 100 nm. In addition, they also displayed a broad dispersity in morphologies, including spheres, cylinders and branched structures. A cluster with core diameter 30 nm, such as that pointed by the arrow contains typically 20 nanoparticles.[51] This wide variety of colloidal morphologies was already noticed with iron oxide nanoparticles, and identified as a major shortcoming of the direct mixing approach.[22]

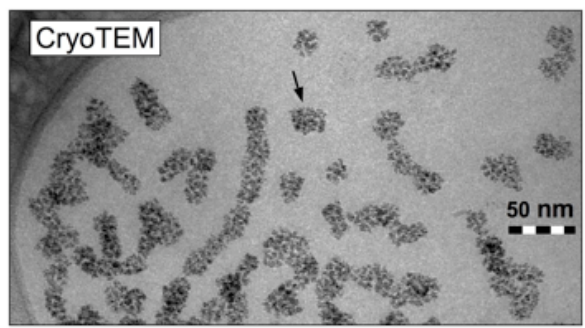

*Figure 3* : Cryogenic transmission electron microscopy (cryo-TEM) images of CeO$_2$-PAA$_{2K}$/PTEA$_{11K}$-b-PAM$_{30K}$ complexes obtained by the direct mixing method. The arrow indicates a cluster with diameter mean 30 nm and aggregation number 20 (number of particles in the three dimensional structure). The image covers a field of 0.45×0.23 μm$^2$.

III.2 - Dialysis

In the three other protocols, particles and polymers were first mixed at high salt concentration. The electrostatic interactions between the oppositely charged species were monitored by a slow and controlled removal of the salt (see Table I for details on initial and final ionic strengths). Fig. 4 illustrates the variation of the hydrodynamic diameter of aggregates obtained by dialysis. Each closed circle there represents a solution that was dialyzed several hours against DI-water, or against with DI-water prepared at a given ionic strength. As a result of dialysis with 10 kD cutoff membrane, an abrupt transition was revealed with decreasing $I_S$, at a critical value $I_S^C$ comprised between 0.4 M and 0.5 M of NH$_4$Cl. This transition is one important findings of the paper. Below the critical ionic strength, large aggregates with hydrodynamic diameters $D_H$ between 300 nm – 500 nm were formed. Data for the individual components, that is the PAA$_{2K}$-coated CeO$_2$ nanoparticles (empty circles) and for PTEA$_{11K}$-b-PAM$_{30K}$ (empty squares) are shown for comparison. For the systems considered separately, the hydrodynamic sizes remained constant over the whole $I_S$-range. The solutions of the individual components were not obtained by dialysis, but prepared by simple mixing. For these systems, since their stability was not affected by the salt, the different methods gave identical results.

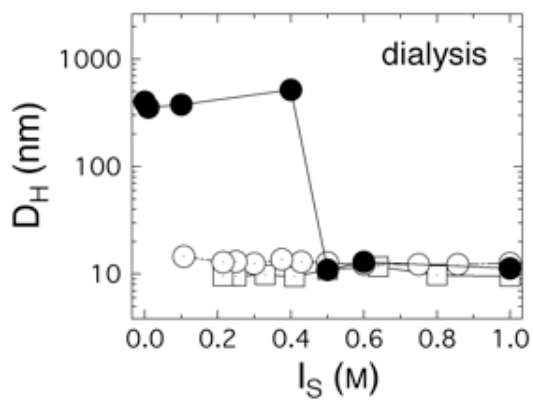

*Figure 4 : Ionic strength dependence of the hydrodynamic diameter for a dispersion containing CeO$_2$-PAA$_{2K}$ particles and oppositely charged PTEA$_{11K}$-b-PAM$_{30K}$ block copolymers (closed symbols). The dispersions were obtained by dialysis. With decreasing $I_S$, an abrupt transition was observed at the critical value comprised between 0.4 M and 0.5 M NH$_4$Cl. The open symbols represent the hydrodynamic diameters for CeO$_2$-PAA$_{2K}$ (circles) and PTEA$_{11K}$-b-PAM$_{30K}$ (squares), respectively.*

Fig. 5 displays a TEM image of nanoparticles aggregates obtained by dialysis. The solution placed under scrutiny here was prepared at a concentration c = 0.2 wt. % and mixing ratio X = 0.5. The dispersion was characterized by an hydrodynamic diameter $D_H$ = 310 nm, with a polydispersity index of 0.2. Examples of aggregates as seen by TEM are shown in Figs. 5a – 5d. In contrast to the direct mixing technique, these clusters exhibit an almost spherical shape. The aggregates were also found to be slightly polydisperse, as illustrated by the TEM-images in Figs. 5a - 5d (diameters between 75 and 220 nm). Note that the diameters obtained by TEM appeared smaller than those determined by dynamic light scattering. The reasons for such discrepancies have been already evaluated[50] : i) the nanoparticle clusters were assumed to be surrounded by a 20 nm-thick polymer corona, made from the neutral block of the copolymer; and ii) when particles are distributed, as it is the case here, light scattering is known to reveal more strongly the largest particles of the distribution, yielding hydrodynamic diameters larger that the average values. The isotropic shape of the nanoceria aggregates suggests that they were grown by a nucleation and growth process, rather than by diffusion- or reaction-limited aggregation which generally yields fractal structures.[52] It should be noted that the aggregates exhibited also a remarkable colloidal stability with time, since neither destabilization nor collapse of the constructs was noted over period of months. This enhanced stability could arise from the presence of a neutral polymer brush (estimated at 20 nm[50]) around the clusters. Assuming a volume fraction of 0.25 inside the large spheres,[22] we have estimated that a 200 nm aggregate was built from ~ 6000 particles.

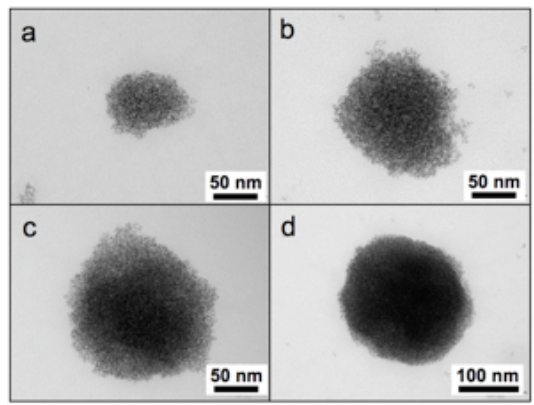

*Figure 5 : TEM images of spherical aggregates obtained by dialysis. The sizes of the aggregates shown are 75 nm (a), 140 nm (b), 160 nm (c) and 220 nm (d). With a volume fraction of 0.20 equal for each aggregates, the aggregation numbers are estimated at 250, 1600, 2400 and 6200 nanoparticles, respectively.*

III.3 - Dilution and Quenching

In the dilution experiments, DI-water was added stepwise to an initial nanoparticle and polymer solutions at 1 M NH$_4$Cl, at the average flow rate of Q = 0.3 μl/s. As for dialysis, the initial dispersion was formulated at c = 0.2 wt. % and X = 0.5. After each step, the scattering intensity and hydrodynamic diameter were determined. Doing so, the ionic strength was also decreased at a rate $dI_S/dt$ ~ $1 \times 10^{-4}$ M s$^{-1}$ comparable with that of the dialysis (Eq. 2). With an overall dilution of a factor 20, the final dispersion had a ionic strength of $5 \times 10^{-2}$ M in ammonium chloride (Table I). Fig. 6a shows the hydrodynamic diameter as a function of the ionic strength of the solution. As for dilution, an abrupt transition was observed at a critical ionic strength estimated here at $I_S^C$ = 0.43 ± 0.01 M. This value was in good agreement with that of dialysis. Moreover, below $I_S^C$, the diameter of the scattering entities stabilized around $D_H$ = 300 - 500 nm, a value close to that of the dialysis. The good superimposition of the $D_H$ versus $I_S$ data for the two techniques suggests that dialysis and slow dilution are equivalent in terms of induced nanostructures.

In the fourth process, the salted dispersions were diluted rapidly and brought to low ionic strength in a time of the order of 1 s. Because the process was fast, it was regarded as equivalent to a quench of the cluster kinetics. In the quenching experiments, $dI_S/dt$ reached values of the order of 1 M s$^{-1}$ (Table I). Under these conditions, nanoparticles aggregates formed spontaneously with sizes in the 100 nm-range, i.e. with values smaller than those obtained by dialysis or slow dilution. These results indicate that the desalting kinetics plays a crucial role in the determination of the cluster sizes.

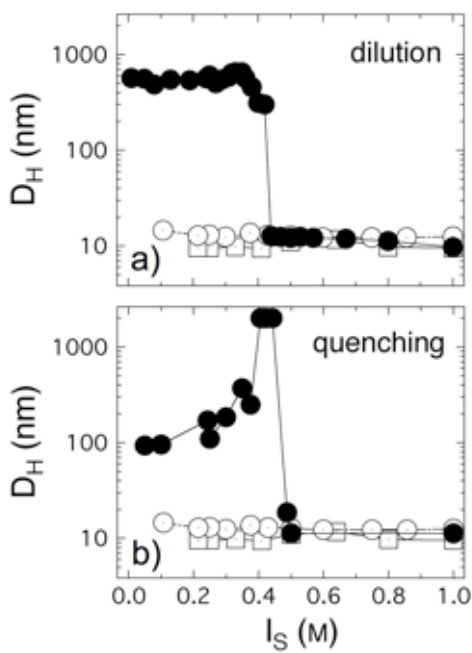

*Figure 6 : Same as in Fig. 4 for dilution (a) and quenching protocols (b). The critical ionic strengths were estimated at $I_S^C$ = 0.43 ± 0.01 M and $I_S^C$ = 0.44 ± 0.02 M, in excellent agreement with that of dialysis.*

Fig. 6b displays the hydrodynamic diameters of dispersions that were quenched at different ionic strengths $I_S^f$ between 0.05 M and 1 M. The main result is here again the observation of a transition between an unassociated and an associate state. This transition occurred at $I_S^C$ = 0.44 ± 0.02 M, in good agreement with the dialysis and dilution processes. Here however, the dispersions were let free to evolve with time after the quench. Between 0.05 M and $I_S^C$, an increase of the diameter $D_H$ followed by the divergence were observed. For final ionic strength comprised between 0.35 M and 0.44 M NH$_4$Cl, the aggregates were beyond the range accessible to light scattering. For practical reasons, they were attributed a value of 2000 nm on the diagram. Between 0.35 M and 0.44 M, the aggregates were micron-sized and sedimented rapidly with time, indicating an instable regime in the cluster growth. In Figs. 4 and 6, the $D_H$'s of the individual components CeO$_2$-PAA$_{2K}$ and PTEA$_{11K}$-b-PAM$_{30K}$ were included for comparison. In conclusion for this part, we have found that the decrease of the ionic strength in the system CeO$_2$-PAA$_{2K}$/PTEA$_{11K}$-b-PAM$_{30K}$ results in an abrupt transition between a fully dispersed state and a clustered state. This transition occurs at a critical value $I_S^C$ = 0.43 M, which is identical for the three desalting pathways. In addition to the strong dependence of the diameter on the desalting kinetics, the existence of stable clusters of different sizes below 0.2 M suggests that the aggregates are kinetically frozen. Once they are formed, they did not evolve, nor destabilize with time. Other pathways in the ionic strength diagram were also attempted, such as step-up increase of the ionic strength or the complexation with copolymers of different molecular weights. The results were consistent with the data shown in Figs. 4 and 6. Note finally that dialysis or dilution experiments made at different X between 0.1 and 1 with the same set of PAA$_{2K}$-coated particles and copolymers yielded very similar results in terms of critical ionic strength and cluster sizes.

Fig. 7 compares the variation of the hydrodynamic diameter during slow dilution for two systems, CeO$_2$-PAA$_{2K}$/PTEA$_{11K}$-b-PAM$_{30K}$ (close circles) and PAA$_{2K}$/PTEA$_{11K}$-b-PAM$_{30K}$ (open circles). The data for the nanoparticles are those of Fig. 6a. The figure provides a comparison of the transition for the particles and for their coating taken separately. For the purely organic system, a transition between unassociated and aggregated state was observed at a slightly lower critical ionic strength, $I_S^C$ = 0.36 ± 0.02 M. The arrow in the figure is associated to a range where two relaxation modes were seen in the autocorrelation functions of the scattered light. The fast mode could be linked to free polymers (*i.e.* with $D_H$ ~ 11 nm) and the slow mode was originating from the PTEA$_{11K}$-b-PAM$_{30K}$/PAA$_{2K}$ clusters. Below 0.2 M, only one mode was observed with a $D_H$-values around 400 nm. These similar behaviors suggest that the transition occurred according to the same scheme for the two systems.

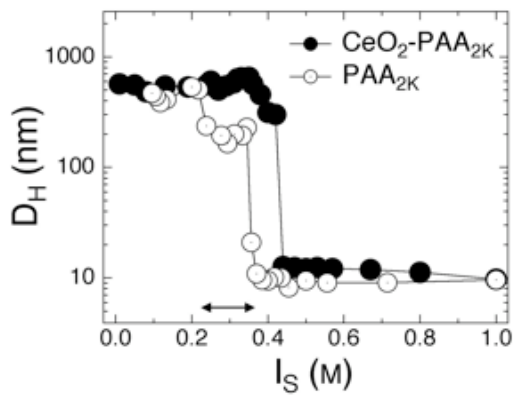

*Figure 7 : Ionic strength $I_S$ dependence of the hydrodynamic diameter $D_H$ for CeO$_2$-PAA$_{2K}$/PTEA$_{11K}$-b-PAM$_{30K}$ and PAA$_{2K}$/PTEA$_{11K}$-b-PAM$_{30K}$ dispersions obtained by slow dilution. For the purely organic system, the critical ionic strength was found at $I_S^C$ = 0.36 ± 0.02 M. The arrow in the figure is associated to a range where both polymers and aggregates were detected by light scattering.*

As already said, this scheme is based on electrostatic interactions between the charged moieties located either along a copolymer backbone, or on the particle surfaces. Here, we show that there is no specificity in the transition (except for the slight reduction of the critical value) that could differentiate polymers dispersed in a solvent or tethered at the surface of particles.

Fig. 8 summarizes the whole set of data obtained on the hydrodynamic diameters by the dilution technique. In Fig.

8, $D_H$ was plotted against the desalting rate $dI_S/dt$ in the range $10^{-4}$ - 1 M s$^{-1}$. For fast dilution or quenching, we observe that the aggregates remained in the 100 nm range. With decreasing $dI_S/dt$, $D_H$ increased and displayed an asymptotic scaling law of the form $D_H \sim (dI_S/dt)^{-1/2}$. The transition between a flat behavior at fast rates and the $(dI_S/dt)^{-1/2}$ scaling at low rates is continuous. Note that the values at low rates coincide with those found by direct mixing, suggesting that in terms of kinetics of complexation, the mixing of oppositely charged species is equivalent to a quench. Also similar are the data obtained dialysis and the slow dilution, for which the aggregates are of the same order, with $D_H$ = 300 - 500 nm. The $D_H \sim (dI_S/dt)^{-1/2}$ scaling observed in the regime of slow kinetics is an interesting result since it could provide some insight about the mechanism of cluster formation. Assuming that the clusters made by dilution are similar to those obtained in a kinetic experiment at comparable times, the above scaling could be transformed into a growth law of the form : $D_H \sim t^{1/2}$. The square root exponent is reminiscent of the late stages of nucleation and growth kinetics (Ostwald ripening) in processes dominated by surface controlled kinetics.[53,54]

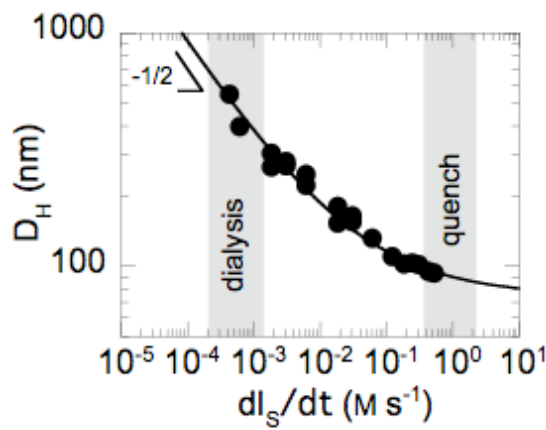

**Figure 8** : *Variation of the cluster diameters $D_H$ as a function of the desalting rate $dI_S/dt$ for slow and rapid (quench) dilution. At low $dI_S/dt$-values, $D_H$ displayed an asymptotic scaling law of the form : $D_H \sim (dI_S/dt)^{-1/2}$.*

## IV - Concluding Remarks

In this work, we have extended our earlier approaches of the electrostatic complexation using oppositely charged particles and polymers. We have considered the system made from PAA$_{2K}$-coated nanoparticles and oppositely charged block copolymers, because these dispersions exhibit remarkable properties, in particular to be stable at high ionic strength. We have borrowed the rationale of the desalting kinetics from molecular biology investigations of chromatin. Described as the lowest hierarchical level of chromosomes, nucleosomes, the basic repeating unit of chromatin fibers, can be reconstituted *in vitro* from its pure components (mainly histones and DNA) by various techniques.[55] This reconstitution process can be seen as a complexation between an excess of polyanions (146 bp of DNA per nucleosome, with a net charge of -294) and a limited pool of polycations (an octamer of basic histones per nucleosome, with a net charge of +146).[56] Since a simple mix of histones and DNA readily produces non-physiologically relevant nucleo-protein aggregates, more gentle methods based on dialysis or step dilution from high salt have been set up and are now the most common methods used to build chromatin fibers relevant for *in vitro* studies.[57,58]

In this paper, we investigate the role of the desalting kinetics on the formation of nanoparticles clusters. This approach was exploited recently to design superparamagnetic nanorods in the micrometer range from 7 nm iron oxide nanoparticles.[51] Here, we define the rate at which the ionic strength is decreased as the control parameter for the kinetics. In typical experiments, the ionic strength is lowered from 1 M to 10$^{-2}$ M NH$_4$Cl, in a time that ranged from 1 second to 1 day. One important result is the observation of an abrupt transition between a dissociated and a clustered state, at a critical ionic strength $I_S^C$ = 0.43 M. Recently, Stoll and coworkers have studied the adsorption properties of a polyelectrolyte on an oppositely charged sphere by Monte Carlo simulations.[4,59,60] This study was performed as a function of several parameters, including the ionic strength. As $I_S$ was lowered in the simulation below a critical value estimated at ~ 0.5 M, these authors have found that the number of monomers adsorbed on the sphere exhibits an abrupt transition between a desorbed and an adsorbed state (see Fig. 6 in Ref.[59]). Similarly, in the desalting process discussed here, the driving force for the association might be related to the desorption-adsorption transition of the polyelectrolyte blocks onto the nanoparticle surfaces.

Concerning the nature of the transition, some results should be emphasized :

   *i)* The sphericity of the CeO$_2$-PAA$_{2K}$/PTEA$_{11K}$-*b*-PAM$_{30K}$ clusters in Fig. 5 suggests that the aggregation results from a nucleation and growth process rather than from a diffusion- or reaction-limited aggregation processes, which generally lead to fractal aggregates;[52]

   *ii)* The quenching experiments performed between 0.35 M and $I_S^C$ show the existence of a macroscopic phase separation;

   *iii)* The critical ionic strength is independent of the desalting rate and process. These findings may indicate that the transition at $I_S^C$ is of first-order character. As the ionic strength is further diminished, a second transition shows up that could be described as a "freezing" transition. This terminology is used in reference to the findings that below 0.2 M the clusters are kinetically frozen. The exact nature of these two transitions should be confirmed.

Even if the variation of the diameter with $dI_S/dt$ in Fig. 8 is continuous, small and large clusters do not share the same exact microstructure. For instance, the 100 nm-clusters obtained by quenching or by direct mixing can be described as core-shell colloids. From the size of the core (Fig. 3), it is suggested that the charged blocks are located in the cores, whereas the neutral poly(acrylamide) blocks are located outside in the corona. This can not be true for clusters with diameters 500 nm. From the geometry of the clusters and the size of the diblocks (~ 100 nm in their fully extended configuration), it should be realized that the neutral chains are incorporated into the cores as well, in an amount that is estimated at 20 % by weight. Although this configuration is plausible, it is not clear so far how this incorporation modifies the formation of the clusters. It is finally important to mention that the present approach is not specific to one system, but could be extended to all types of oppositely charged systems. One condition for its application is the colloidal stability of the dispersions at high salt content

## Appendix

In this section, we model the X-dependence of the scattering intensity obtained for aqueous dispersions containing oppositely charged polymers and nanoparticles and obtained by direct mixing[35]. The model explicitly assumes that the mixed aggregates are formed at a fixed polymer-to-nanoparticle ratio, regardless of X. The model was called stoichiometric for this reason.

In the direct mixing formulation, the total concentration of active matter is kept constant. The respective nanoparticle and polymer concentrations in the mixed solutions vary as a function of X as :

$$c^0_{Pol} = \frac{c}{1+X} \; ; c^0_{Part} = \frac{cX}{1+X} \quad \text{(A-1)}$$

The exponent "0" in Eq. A-1 refers to the concentrations of all polymers and particles, present in a complexed or in an uncomplexed state. The present model is based on the hypothesis that the solutions contain stoichiometric polymer/particle aggregates at all values of X. The ratio noted r, between polymers and nanoparticles may be expressed as a function of the preferred mixing ratio $X_P$ as :

$$r = \frac{1}{X_P} \frac{M^{part}_w}{M^{pol}_w} \quad \text{(A-2)}$$

Eq. A-2, together with the experimental determination of $X_P$ were utilized in our previous work to obtain an estimate of r[34,35,50]. Under the assumption of fixed r, $X < X_P$ corresponds then to the domain where the polymers are the major component and where all added nanoparticles participate to the aggregates. For $X > X_P$, it is the reverse : the nanoparticles are in excess and all polymers are used to form the mixed colloids. In the first regime, mixed aggregates are in equilibrium with free polymers, and in the second with free particles. Therefore, a mixed solution prepared at X may comprise all three species. The scattering intensity expresses as the sum of the three contributions :

$$\mathcal{R}(q,c,X) = \sum_i K_i c_i(X) \left[ \frac{1}{M^i_w}\left(1 + \frac{q^2 R^2_{G,i}}{3}\right) + 2A^i_2 c \right]^{-1} \quad \text{(A-3)}$$

where the index i refers to polymers, mixed aggregates and particles. In Eq. A-3, $M_{w,app}$ is the apparent molecular weight, $R_G$ the radius of the gyration, $K = 4\pi^2 n^2 (dn/dc)^2/\mathcal{N}_A \lambda^4$ the scattering contrast coefficient ($\mathcal{N}_A$ is the Avogadro number) and $A_2$ is the second virial coefficient, dn/dc the refractive index increments and q the wave-vector at which the light scattering is made. For concentration and wave-vector extrapolated to 0, Eq. A-3 simplifies in :

$$\mathcal{R}(q \to 0, c, X) = K_{Pol} M^{Pol}_w c_{Pol}(X) + K_{Agg} M^{Agg}_{w,app} c_{Agg}(X) + K_{Part} M^{Part}_w c_{Part}(X) \quad \text{(A-4)}$$

where the equality $\sum_i c_i(X) = c^0_{Pol} + c^0_{Part}$ insures the mass conservation of the different species. In the following, Eq. A-4 will be assumed to describe the intensity measured in the different systems at c = 0.1 wt.% and θ = 90°. For aggregates with a radius of gyration $R_G$ ~ 30 nm, the term $\frac{1}{3} q^2 R^2_G$ in Eq. A-3 represents only 15 % of the total scattered intensity. For larger aggregates, the q-dependence of the intensity has to be taken into account.

At low X, the Rayleigh ratio normalized to its value at X = ∞ ($\mathcal{R}_\infty = K_{Part} M^{Part}_w c$) expresses as :

$$\mathcal{R}'(X < X_P) = K'_{Pol} m \frac{X_P - X}{X_P(1+X)} + K'_{Agg} M \frac{X(1+X_P)}{X_P(1+X)} \quad \text{(A-5a)}$$

The first contribution arises from the unassociated polymers and the second term from the mixed aggregates. Similarly, at large X, the coexistence occurs between the particles and the aggregates and the intensity goes as :

$$\mathcal{R}'(X > X_P) = K'_{Agg} M \frac{1+X_P}{1+X} + \frac{X - X_P}{1+X} \quad \text{(A-5b)}$$

where the notations

$$K'_{Pol} = \frac{K_{Pol}}{K_{Part}}, \; K'_{Agg} = \frac{K_{Agg}}{K_{Part}},$$

$$m = \frac{M^{Pol}_w}{M^{Part}_w}, \; M = \frac{M^{Agg}_{w,app}}{M^{Part}_w} \quad \text{(A-6)}$$

have been adopted. With these notations, Eq. A-2 may be rewritten $r = (mX_P)^{-1}$. It is interesting to note that the normalized scattered intensity in Eqs. A-5 does not depend on the total concentration c. The expression should hence be valid at all c in the dilute regime. Eqs. A-5 were used to fit the scattering data of Fig. 2a keeping r and M as adjustable parameters. All others quantities, such as the coupling constants $K'_i$ and molecular weights of single constituents were known. The values for dn/dc were estimated from the weighted sum of the increments of each component. For the calculations of Eqs. A-5, we have taken $K'_{Pol} = 0.9$, $K'_{Agg} = 1$, m = 0.0814 based on previous studies[35,61], and we have found the complexation occurred with r = 8 polymers per particle. The clusters have average molecular with M = 30 times those of the bare particles. With the approximations made, this result is in relative good agreement with the cryo-TEM image of the clusters (Fig. 3). Note in Fig. 2 that the agreement between the model and the data is poor apart from the scattering peak, indicating that the stoichiometric model might not be appropriate in this range [35]. More sophisticated treatments would have required to perform static light scattering experiments as a function of the wave-vector, at all mixing ratio X and at several concentrations for each of these X. The results of the fitting are shown in Fig. 2a as solid curves. Also shown in the figure is the scattering intensity corresponding to the state where particles and polymers remain unaggregated (dashed lines):

$$\mathcal{R}'_{\text{UnAgg}}(X) = \frac{K'_{Pol}m + X}{1 + X} \qquad (A-7)$$

## Acknowledgements

We are grateful to Ling Qi, Jean-Paul Chapel, Jean-Christophe Castaing from the Complex Fluids Laboratory (CRTB Rhodia, Bristol, Pa) and Olivier Sandre from the Laboratoire Liquides Ioniques et Interfaces Chargées (UPMC, Université Paris 6) for discussions and comments during the course of this study; to Eric Le Cam from the Institut Gustave Roussy (Université Paris-Sud) for complementary TEM experiments and helpful discussions. Aude Michel from the PECSA-UPMC-Université Paris 6 is kindly acknowledged for the TEM experiments. This research is supported in part by Rhodia and by the Agence Nationale de la Recherche, under the contract BLAN07-3_206866.